\documentstyle[aps,epsfig]{revtex}

\begin{document}

\title{Vibrational spectra in glasses}

\author{T.~S.~Grigera$^1$, V.~Mart\'{\i}n-Mayor$^1$, G.~Parisi$^1$ and
P.~Verrocchio$^2$}

\address{$^1$ Dipartimento di Fisica, Universit\`a di Roma ``La
Sapienza'', Piazzale Aldo Moro 2, 00185 Roma, Italy \\ INFN sezione di
Roma - INFM unit\`a di Roma \\ $^2$ Dipartimento di Fisica, Universit\`a
di Trento, Via Sommarive, 14, 38050 Povo, Trento, Italy \\ INFM unit\`a
di Trento}

\date{\today}
\maketitle
\begin{abstract}
The findings of X-ray and neutron scattering experiments on amorphous
systems are interpreted within the framework of the theory of
Euclidean random matrices.  This allows to take into account the
topological nature of the disorder, a key ingredient which strongly
affects the vibrational spectra of those systems.  We present a
resummation scheme for a perturbative expansion in the inverse
particle density, allowing an accurate analytical computation of the
dynamical structure factor within the range of densities encountered
in real systems.
\end{abstract}

\section{Introduction}


Propagating density fluctuation of macroscopic size (hydrodynamic
limit) are known to exist both in ordered and in disordered materials.
Whereas in the ordered ones those excitations (phonons) persist up to
momenta of about one tenth of the Debye momentum, the fate of
excitations of microscopical size in disordered systems is still a
quite puzzling issue, both from the theoretical and the experimental
point of view.


Recently, a lot of experimental attention (Buchenau {\em et al.} 1986,
Benassi {\em et al.} 1996, Foret {\em et al.} 1996, Masciovecchio {\em
et al.} 1996, Masciovecchio {\em et al.} 1997, Masciovecchio {\em et
al.} 1998, Monaco {\em et al.} 1998, Sette {\em et al.} 1998, Fioretto
{\em et al.} 1999, Ruocco {\em et al.} 1999, Sokolov {\em et al.}
1999) has been paid to the high frequency dynamics of disordered
systems like glasses.  As a matter of fact, high-resolution inelastic
X-ray scattering (IXS) and neutron scattering techniques have made
accessible to experiment the region where the exchanged external
momentum $p$ is comparable to $p_0$, namely the momentum where the
static structure factor has its first maximum:
\begin{equation} 
0.1 \: p_0 < p < p_0
\label{RANGE}
\end{equation}
The typical momentum $p_0$ can be thought as the generalization of the
Debye momentum to the case of disordered systems.  With these data
available for the study of excitations of microscopic size, one is
naturally led to ask which of the well-known features of phonons
survive even in disordered systems.  A number of facts have emerged
from the experiments:
\begin{enumerate}
\item The dynamic structure factor $S(p,\omega)$ has a Brillouin-like
peak for momenta up to $p/p_0 \sim 0.5$.  This inelastic peak is due
to the interaction of the external photon (or neutron) with some
excitation of the system.  A very controversial issue is the
propagating nature of the excitations (see, for example, Caponi {\em et
al.} and Foret {\em et al.} in this same volume) within that range of
momenta.  As a guideline, one could say that as long as the dispersion
relation between the position of the peak (in the frequency domain)
and the external momentum $\omega_P (p)$ is nearly linear, the excitations
are likely propagating. Of course, the study of Anderson localization
(see, for instance, Parisi 1994)
shows that, strictly speaking, this condition is neither necessary nor
sufficient, so care should be taken.
\item The peak has a width $\Gamma$, whose dependence on the momentum
$p$ within the range~(\ref{RANGE}) has been described by means of the
scaling law (for a large variety of materials):
\begin{equation}
\Gamma \propto p^{\alpha}
\label{BROAD}
\end{equation}
Let us remark that, while there seems to be a quite general agreement
about the fact that $\Gamma$ is not affected by changes in the
temperature, there is still some debate about the actual value of the
exponent $\alpha$. More precisely, at low enough momenta (e.g. light
scattering) $\alpha$ is undoubtedly $\sim 2$, while at higher momenta
(e.g. X-ray scattering) neither the value $\alpha \sim 2$,
nor the value $\alpha \sim 4$, describing instead a Rayleigh
scattering regime, can be ruled out on experimental grounds.

\item The density of states (DOS) exhibits an excess
respect to the Debye behavior ($g(\omega)\propto\omega^2$), known as
the {\em Boson peak}. This feature is particularly remarkable for
strong glasses. Generally speaking, the typical frequencies of
the Boson peak are in the region where the
relation between the frequency of the Brillouin-peak $\omega_p$ and
the momentum $p$ is still linear.

\item A secondary peak at frequencies smaller than the Brillouin one
develops for large momenta, becoming dominant for $p/p_0 \sim 0.5$.
\end{enumerate} 

\begin{figure}
\epsfig{file=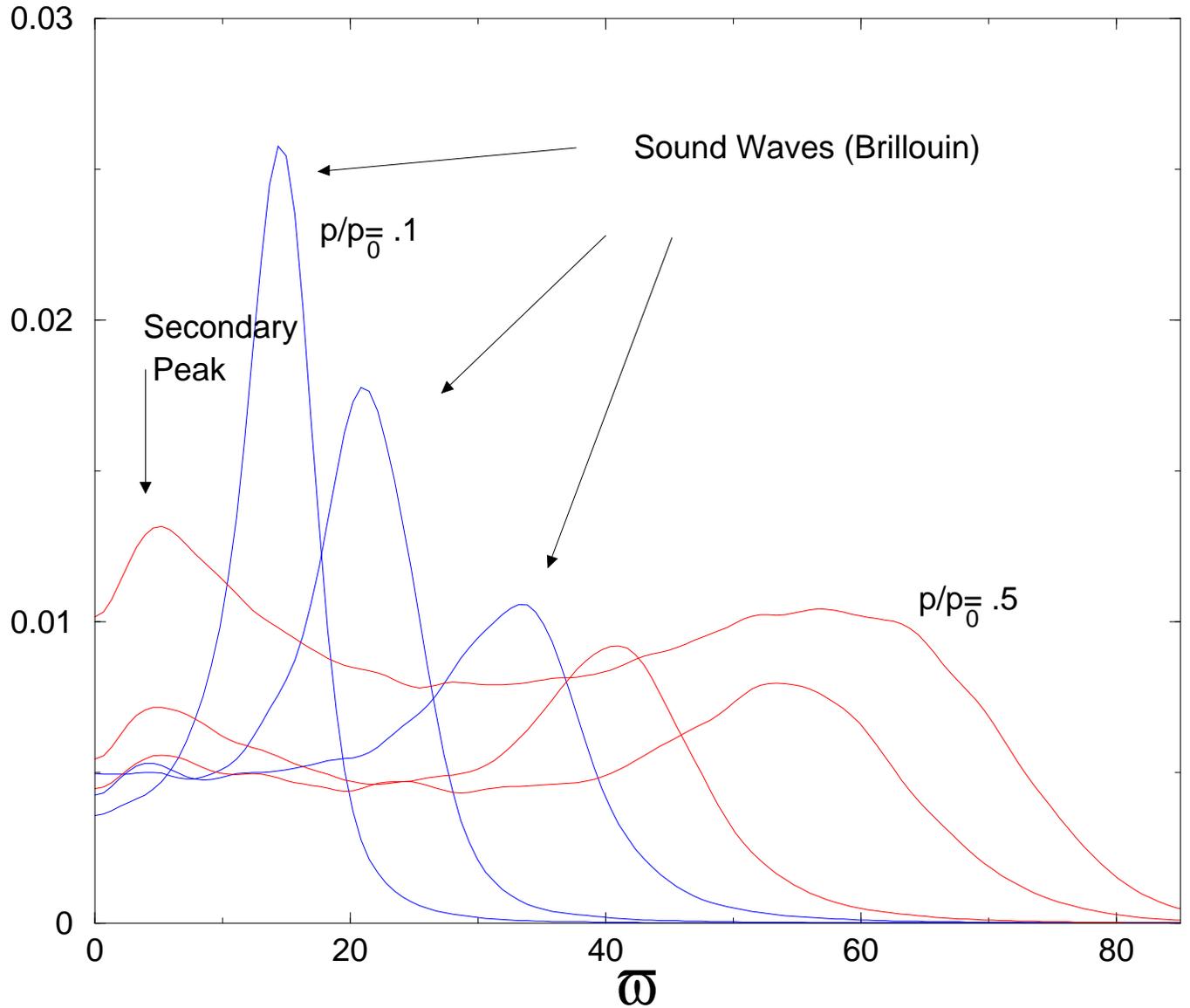, angle=270, width=\columnwidth}
\caption{Dynamical structure factor for argon, obtained numerically by
Ruocco {\em et al.} 2000 at low temperature $\sim 10 K$. The Brillouin-like and the secondary peak appear quite clearly.}
\label{argonruocco}
\end{figure}

A number of basic insights on the spectral properties of glasses have
been obtained by means of molecular dynamics simulations (Mazzacurati
{\em et al.} 1996, Dell'Anna {\em et al.} 1998, Horbach {\em et al.}
1998, Ribeiro {\em et al.} 1998, Sampoli {\em et al.} 1998, Allend
{\em et al.} 1999, Feldman {\em et al.} 1999, Taraskin and Elliot
1999) on systems such as argon, silica and water.  Among the others,
let us point out the fact that, in the glass phase, the high frequency
dynamics is very well described in the framework of the {\em harmonic
approximation}. Both dynamic quantities like the dynamic structure
factor (Ruocco {\em et al.} 2000) and thermodynamic observables as the
specific heat (Horbach at al 1999) have been shown to be correctly
described by models where only vibrations around quenched positions
are taken into account.  In figure (\ref{argonruocco}) the numerical
results obtained by Ruocco {\em et al.} (2000) are shown, which
illustrate the above-mentioned spectral features.


On the theoretical side, the study of the short time
properties of glasses has been attempted from two different (and
somewhat complementary) points of view.

\begin{itemize}
\item On one hand, it has been shown (G\"otze and Mayr 2000) that the
Mode Coupling Theory (MCT), which usually describes the long time
limit of the time correlators, can be modified in the glassy phase so
as to describe only the excitations around the quenched structure.
Within such an approach, $S(p,\omega)$ was computed for a hard-sphere
glass by means of a {\em generalized hydrodynamics approximation}.  In
this framework, a Brillouin peak with a linear dispersion relation
({\em propagating excitations}) is obtained up to $p/p_0 \sim 0.5$, as
well as a secondary peak for higher values.  Interestingly, the peak
width $\Gamma$ was shown to follow the simple scaling law:
\begin{equation}
\Gamma_{MCT} \propto p^2
\end{equation}
only in the very low momenta regime $p/p_0 < 0.1$, whereas a different
law applies to the range of momenta comparable with the experimental
and simulation values (see the figure (8) in G\"otze and Mayr 2000)
which seems hard to reduce to the simple form~(\ref{BROAD}).
Moreover, it has been shown (Ruocco {\em et al.} 2000) that, resorting
to the harmonic approximation, the memory function can be simply
obtained by means of the eigenvalues and eigenvectors of the Hessian
matrix.

\item A second approach relies on the study of the statistical
properties of random matrices (Metha 1991), since within the harmonic
approximation the whole dynamical features are encoded in the system's
Hessian matrix.  A crucial point is to distinguish two different
classes of systems:
\begin{enumerate}
\item vibrational systems whose disorder has a topological origin, as
in glasses (Elliot 1983)
\item vibrational systems on a lattice, where random couplings
constants (Schirmacher {\em et al.} 1998, Montagna {\em et al.} 1999,
Kantelhardt {\em et al.} 2000, Martin-Mayor {\em et al.} 2000,
Taraskin {\em et al.} 2001) are introduced in order to mimic the
behavior of real glasses.
\end{enumerate}
The latter class has recently been studied in some detail (Schirmacher
{\em et al.} 1998, Taraskin {\em et al.} 2001) in order to give an
insight into the spectral properties, in particular into the Boson
Peak.  It turns out (Martin-Mayor {\em et al.} 2000), however, that
these models definitely miss the $p^2$ behavior of the peak width.
Because of the long range order due to the underlying lattice, the
general beahviour is given instead by
\begin{equation}
\Gamma_{CPA} \propto p^4
\end{equation}
even at very low momenta.  The above result can be easily obtained
within the CPA approximation and can be checked by the direct
inspection of the eigenvectors.  The discrepancy with the low momenta
behaviour of glasses being due precisely to the lack of topological
disorder in lattice models, it seems reasonable to address the study
of the former class.  The problem turns out to reduce to the study of
a very special class of random matrices, which have been called {\em
Euclidean Random Matrices} (ERM) (Mezard {\em et al.} 1999).  Is is
worthwhile to note that this approach would allow to address, besides
the study of the high frequency regime of glasses, even the problem of
instantaneous normal modes (INM) i.e. the statistical properties of
the Hessian matrix of a liquid at equilibrium (Wu and Loring 1992, Wan
and Stratt, 1994, Keyes 1997, Biroli and Monasson 1999), the framework
being very similar (Cavagna {\em et al.} 1999).  Nevertheless, in the
following we shall focus only on the latter.  Recently (Martin-Mayor
{\em et al.} 2001), the dynamical structure factor $S(p,\omega)$ was
analytically computed.  The computation relies on a pertubative
expansion, the expansion parameter being the inverse density particle
$1/\rho$.  Let us recall what those pertubative results undoubtedly
have shown:
\begin{itemize}
\item At $\rho \to \infty$, the dynamical structure factor is composed
of a single delta function, whose position changes linearly with the
external momentum $p$ at low momenta, representing the undamped
propagation of a sound wave in an elastic medium.  This can be easily
understood by considering the infinite number of particles per
wavelength in that limit.
\item For high but finite densities, the disorder of the position of
the particles involved in the propagation causes a broadening of the
Brillouin peak. In other words, the plane waves are no longer
eigenstates of the Hessian, hence a finite spreading of the
eigenfrequencies involved arises.  In principle no analogy with the
Rayleigh scattering should be expected. As a matter of fact, the
perturbative computation, performed up to $1/\rho^2$ order, shows that
at low enough momenta the general behaviour is instead given by:
\begin{equation}
\Gamma_{ERM} \propto (1/\rho)^2 p^2.
\end{equation} 
Yet at density $\rho=1$ a significant deviation from that behaviour is
seen as $p/p_0$ becomes greater than $\sim 0.1$, in a good qualitative
agreement with the results of the MCT approach.
\end{itemize}
\end{itemize}


The perturbative calculation has the problem that both the structure
factor and DOS end abruptly at a cut-off frequency.  Moreover, the
experimental densities are of order $\rho \sim 1$ or less (in reduced
units), hence the ability of perturbation theory to catch the features
of real systems is at least questionable.
 
In order to overcome those difficulties, we present a resummation
scheme for the perturbative results (Grigera {\em et al.} 2001) which
includes terms at all orders in $1/\rho$.  Within this scheme, a very
good agreement with numerical results, even at large frequencies, is
found for values of $\rho > 0.3$ (reduced units).

\section{harmonic approximation}


We shall assume that the particles of our system in the glass phase
can only oscillate around their equilibrium position, claiming that
this is enough to predict the high frequency properties.  Here by
``equilibrium'' we do not mean thermodynamic equilibrium (the
thermodynamics of glasses is an entirely different issue than the one
addressed here), but rather mechanical equilibrium (i.e. a position
where the forces on all particles are zero).  For the sake of
simplicity, let us consider displacements only along a given direction
${\mbox{\boldmath$u$}}$:
\begin{equation}
{\mbox{\boldmath$x$}}_j(t)={\mbox{\boldmath$x$}}_j^{\mathrm{eq}}+
{\mbox{\boldmath$u$}}\varphi_j(t) 
\end{equation}

Since the number of equilibrium positions
$\{{\mbox{\boldmath$x$}}_j^{\mathrm{eq}}\}$ available to the system is
actually infinite, growing exponentially with the number of particles
$N$, one can ask which one is to take in order to compute the
vibrational spectra.  The answer lies on the so called {\em self
averaging} hypothesis, which has proved to be correct for many
observables in disordered systems.  Hence, we shall assume that {\em
in the thermodynamic limit}, the spectra obtained for the different
realizations of disorder is equal to the one computed by considering
the average over the disorder.  This statistical approach leads to a
tremendous simplification of our task, as shown in the following,
because the model is defined simply by the probability distribution of
the random variables $\{{\mbox{\boldmath$x$}}_j^{\mathrm{eq}}\} ,\,
j=1,\ldots,N$.

Within that harmonic framework, the energy of the system is ($m$ is
the mass of the particles and $\Omega$ is a frequency scale)
\begin{equation}
V(\{\varphi_i\})=\frac{m \Omega^2}{2}\sum_{i,j} 
f({\mbox{\boldmath$x$}}_i^{\mathrm{eq}}-{\mbox{\boldmath$x$}}_j^{\mathrm{eq}})
\,(\varphi_i-\varphi_j)^2\,.
\label{ENERGIA}
\end{equation}
We choose units such that $m=1$ and $\Omega=1$. Thus the Hessian
matrix
\begin{equation}
M_{ij}=\delta_{ij} \sum_{k=1}^N
f({\mbox{\boldmath$x$}}_i^{\mathrm{eq}}-{\mbox{\boldmath$x$}}_k^{\mathrm{eq}})
-
f({\mbox{\boldmath$x$}}_i^{\mathrm{eq}}-{\mbox{\boldmath$x$}}_j^{\mathrm{eq}})
\label{HESSIANO}
\end{equation}
is an Euclidean random matrix, whose spectral properties we are
studying.

In particular, we are interested both in the dynamic structure factor
and in the density of states (DOS).
\begin{itemize}
\item The dynamic structure factor $S(p,\omega)$, roughly speaking,
gives the spectrum of states ``excited'' by a plane wave with a given
momentum $p$. It can be obtained from the Hessian matrix of the
potential (\ref{ENERGIA}) in the classical limit and in the one
excitation approximation (see for instance Martin-Mayor {\em et al.}
2001 for a detailed discussion). It reads
\begin{equation} 
S^{(1)}(p,\omega)= \frac{k_{\mathrm B} T p^2}{\omega^2}
\overline{\sum_n |\langle n| {\mbox{\boldmath{$p$}}}\rangle|^2
\delta(\omega -\omega_n)}\,
\end{equation}
where $|n\rangle$ are the eigenvectors of the Hessian matrix 
(\ref{HESSIANO}), the eigenfrequencies $\omega_n$ are the square root of the 
eigenvalues (which are all positive, see equation(\ref{ENERGIA})),
$T$ is the temperature, the overline stands for the average on the equilibrium
positions $\{{\mbox{\boldmath$x$}}_j^{\mathrm{eq}}\}$, while
$|{\mbox{\boldmath$p$}}\rangle$ stands for a momentum plane wave
($\langle j| {\mbox{\boldmath$p$}}\rangle= \exp[{\mathrm i}
{\mbox{\boldmath$p$}}\cdot{\mbox{\boldmath$x$}}_j^{eq}]/\sqrt{N}$, where
$|j\rangle$ represents the vector where the $j$-th particle has 
displacement $\mbox{\boldmath$u$}$ and the other particles do not move).


\item The DOS describes instead the density of 'all' the vibrational
states existing within the system.  Interestingly, at the level of
one-excitation approximation, the DOS can be obtained by:
\begin{equation}
g(\omega)=
\frac{\omega^2}{k_{\mathrm B} T p^2}
S^{(1)}(p\to\infty,\omega)
\label{INFINITO}
\end{equation}
The above theoretical result has been pointed out only very recently
(Martin-Mayor {\em et al.} 2001).  Although it can be checked very easily by
means of numerical simulations, we believe it would be also very
interesting to check it on real systems, using the experimental data
at large momenta which are now available.  On the other hand, at very
high frequencies the one-excitation approximation does not hold, and
many-excitations contributions should be taken into account, hence the
relation~(\ref{INFINITO}) is obviously only an approximation, whose
reliability in describing real systems is a very interesting matter.
Moreover, we shall see in the following that~(\ref{INFINITO}) is
crucial in obtaining a model independent derivation of the actual
small $p$ behaviour of the width of the Brillouin peak, holding only
for topologically disordered systems.  As a matter of fact, the same
argument cannot be applied to lattice models, because the
relation~(\ref{INFINITO}) is not true in topologically ordered models.

\end{itemize}


As mentioned, the Hessian is averaged over the disorder, i.e. over all
the allowed equilibrium positions. Those are clearly distributed in a
highly correlated manner, due to the hard-core repulsion and long
range attraction of the potential. It was shown (Martin-Mayor et
al. 2001) that, at the level of the superposition approximation, the
correlations can be taken into account by doing an uncorrelated
average while at the same time suitably renormalizing the interaction
$f$: if the $k$ point static correlation function is written as
\begin{equation}
g(r_1, \dots r_k) = g(r_1) \dots g(r_k),
\end{equation}
then the inverse-density expansion for the dynamic structure factor of
a system with correlation function $g(r)$ and force $F(r)$ can be
obtained from the expansion of a fully uncorrelated system defining a
{\em dressed} interaction:
\begin{equation} 
f(r) \equiv g(r)F(r)
\end{equation}
where the value of the ``spring'' constant $F(r)$ is weighted by the
probability to find two centers of oscillations at the relative
distance $r$.

\section{Non perturbative results}


Let $\hat f(p)$ be the Fourier transform of $f(r)$.  The main 'object'
to compute is the resolvent of the Hessian matrix (\ref{HESSIANO}),
which depends on the complex variable $z = \omega^2 +{\mathrm i}\eta$.
It can be written in the following way:
\begin{eqnarray}
G(p,z)&=&\frac{1}{z-\epsilon(p)-\Sigma(p,z)}\,,\\
\epsilon(p)&=&\rho[\hat f(0)-\hat f(p)]\,
\label{LAMBDADIP}.
\end{eqnarray}
where complex self-energy $\Sigma(p,z)$ has been introduced.
Exploiting the well known relation between resolvent and dynamical
structure factor (letting aside the prefactor $k_{\mathrm B} T p^2$,
unessential for the $\omega$ dependence): 
\begin{equation} 
S^{(1)}(p,\omega)=- {1 \over \omega \pi} \lim_{\eta \to 0}
{\mathrm{Im}}\, G(p,\omega^2 + i \eta).
\end{equation}
the following connections between the main features of the dynamical
structure factor and the self-energy are established:
\begin{itemize}
\item
The 'bare' dispersion relation $\epsilon(p)$, which would give the
position of the peak in the elastic medium limit, is renormalized by
the real part of the self-energy $\Sigma'(p,z)$.  This gives
$\omega^{renorm}(p)$, the position of the maximum of the structure
factor in the frequency domain.  Let us note that $\omega^{renorm}(p)$
is certainly linear for small $p$, as expected.
\item The imaginary part $\Sigma''(p,z)$ 
computed at the position of the peak $\omega=\omega^{renorm}(p)$
gives the width, $\Gamma(p)$, of the $S^{(1)}(p,\omega)$ by means of:
\begin{equation}
\Sigma''(p,\omega^{renorm}(p)) = \omega^{renorm}(p) \Gamma(p)
\label{GAMMA}
\end{equation}
\end{itemize}

\begin{figure}
\epsfig{file=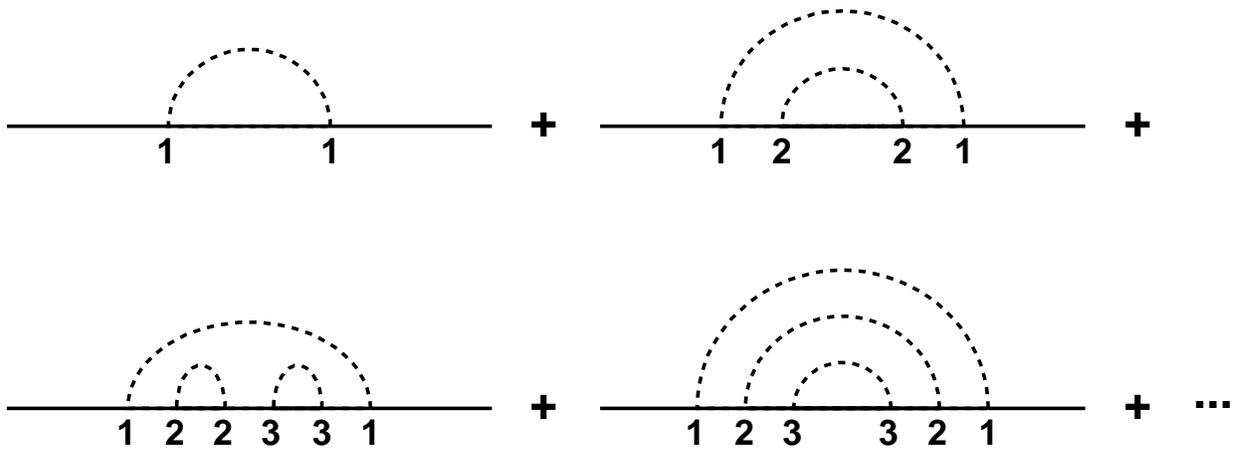, angle=0, width=\columnwidth}
\caption{Cactus diagrams of the $1/\rho$ expansion. The numbers
correspond to the particle-label repetitions}
\label{DIAGRAMMI}
\end{figure}

The self-energy can be obtained as a series in $1/\rho$: the
$k^{\mathrm{th}}$-order corresponds to $k$ particle-label repetitions
when calculating the moments of the $S^{(1)}(p,\omega)$ (the details
can be found in Martin-Mayor {\em et al.} 2001).  It is easy to show
that the sum of all the {\em cactus diagrams} (see figure
(\ref{DIAGRAMMI})) is given by the solution of the following integral
equation:
\begin{equation}
\Sigma(p,z)=\frac{1}{\rho}\int\frac{d^D q}{(2\pi)^D} 
\frac{ \left[\rho\left(\hat f(\mbox{\boldmath$q$})-\hat f(\mbox{\boldmath$p$}-
\mbox{\boldmath$q$})\right)\right]^2}
{z-\epsilon(q)-\Sigma(q,z)}\,.
\label{CACTUS}
\end{equation}

	
Interestingly, Equation(\ref{CACTUS}) provides a model-independent
derivation of the $p^2$ scaling of width of the peak of the
$S^{(1)}(p,\omega)$.  Indeed, the large $q$ contribution to the
imaginary part of the integral in equation (\ref{CACTUS}) is, because
of (\ref{INFINITO})
\begin{equation}
\Sigma''_\infty(p,z)= -\pi \rho g_\lambda(\lambda) \int\frac{d^D q}{(2\pi)^D}
\!\!  \left(\hat f(\mbox{\boldmath$q$})-\hat f(\mbox{\boldmath$p$}-
\mbox{\boldmath$q$})\right)^2.
\label{sigma-infinito}
\end{equation}
where $g_\lambda(\lambda)$ is the density of states in the domain of
eigenvalues ($\lambda=\omega^2,
g_\lambda(\omega^2)=\frac{g(\omega)}{2\omega}$).  If the spectrum is
Debye-like we have $g_\lambda(\lambda) \propto \lambda^{0.5}$, and it is
straightforward to show that (\ref{sigma-infinito}) is proportional to
$\omega^{renorm}(p) \, p^2$. Then the relation (\ref{GAMMA}) implies
the scaling:
\begin{equation}
\Gamma(p) \propto p^2 
\end{equation}
irrespective of the function $f(r)$.  Clearly this is only the large
$q$ contribution to the integral, but it has been shown (Grigera {\em
et al.} 2001) that it indeed controls the peak width at small $p$.

\section{the Gaussian case}

\begin{figure}
\epsfig{file=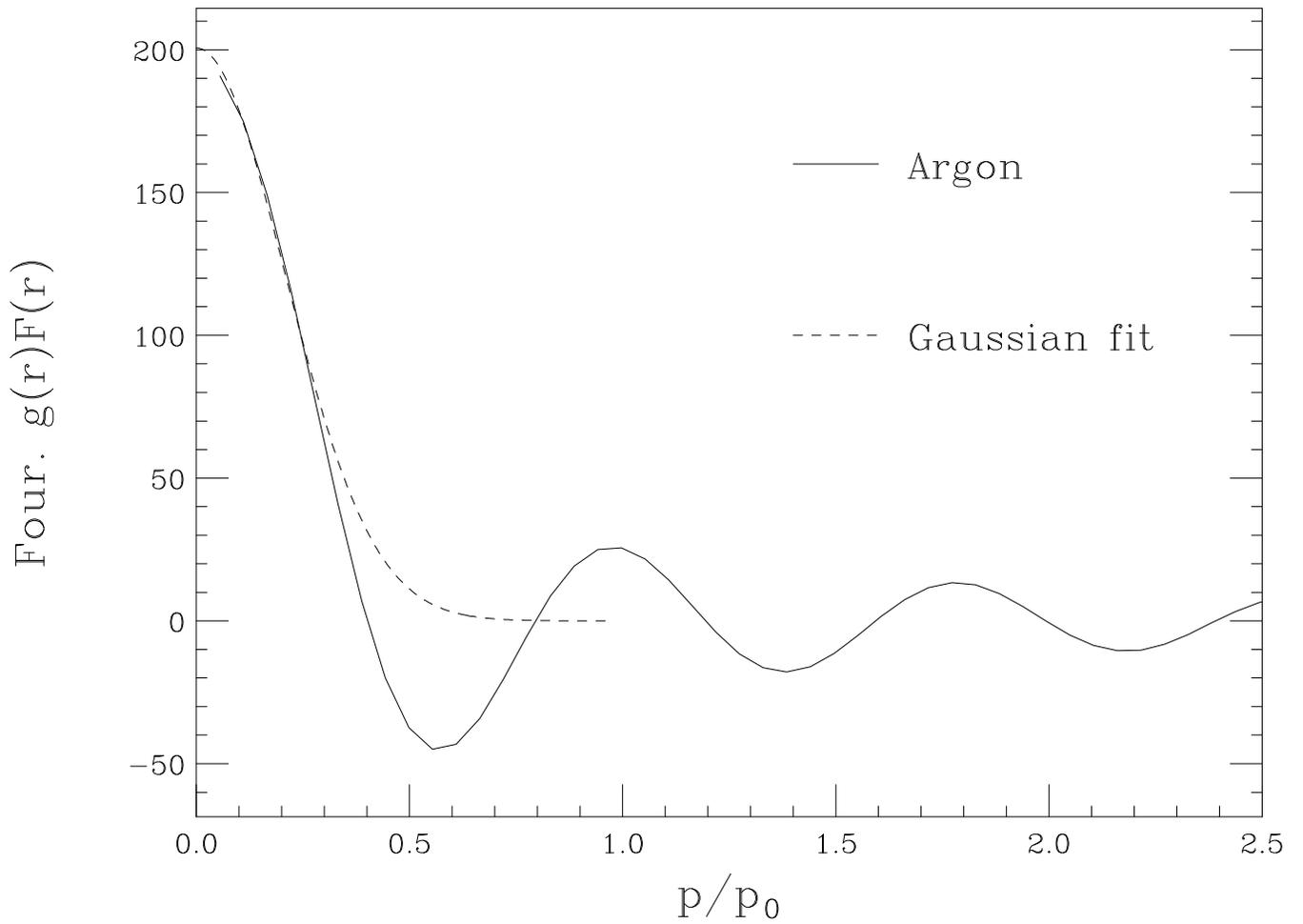, angle=90, width=\columnwidth}
\caption{The gaussian choice for $f(q)$ compared with the Fourier
trasform of $g(r)F(r)$ at $\sim 10 K$, $F(r)$ being the second
derivative of the Lennard-Jones pair potential which is supposed to 
modelize the Argon pair interactions.}
\label{CONFRONTO}
\end{figure}

In the following we shall study the simple case
\begin{equation}
f(r)=\exp[-r^2/(2\sigma^2)].
\label{GAUSSIANA} 
\end{equation}
This may seem an oversimplification, too distant from any realistic
case. However, it is not actually the case, at least for small
momenta. For the sake of comparison we can see in figure
(\ref{CONFRONTO}) the Fourier transform of the function $g(r)F(r)$ for
Argon at very low temperature ($\sim 10 K$) together with that of
(\ref{GAUSSIANA}).  We thus expect our results to be reasonable for
momenta smaller than the first zero of the Fourier transform of
$g(r)F(r)$, which in general is close to the maximum of the static
structure factor, $p_0$.  Since the Fourier transform of our force
decreases an order of magnitude by $p_0=2/\sigma$, we shall take this
as {\em our} $p_0$ during the following discussion, and $\sigma$ will
be our unit of length.

\begin{figure}
\epsfig{file=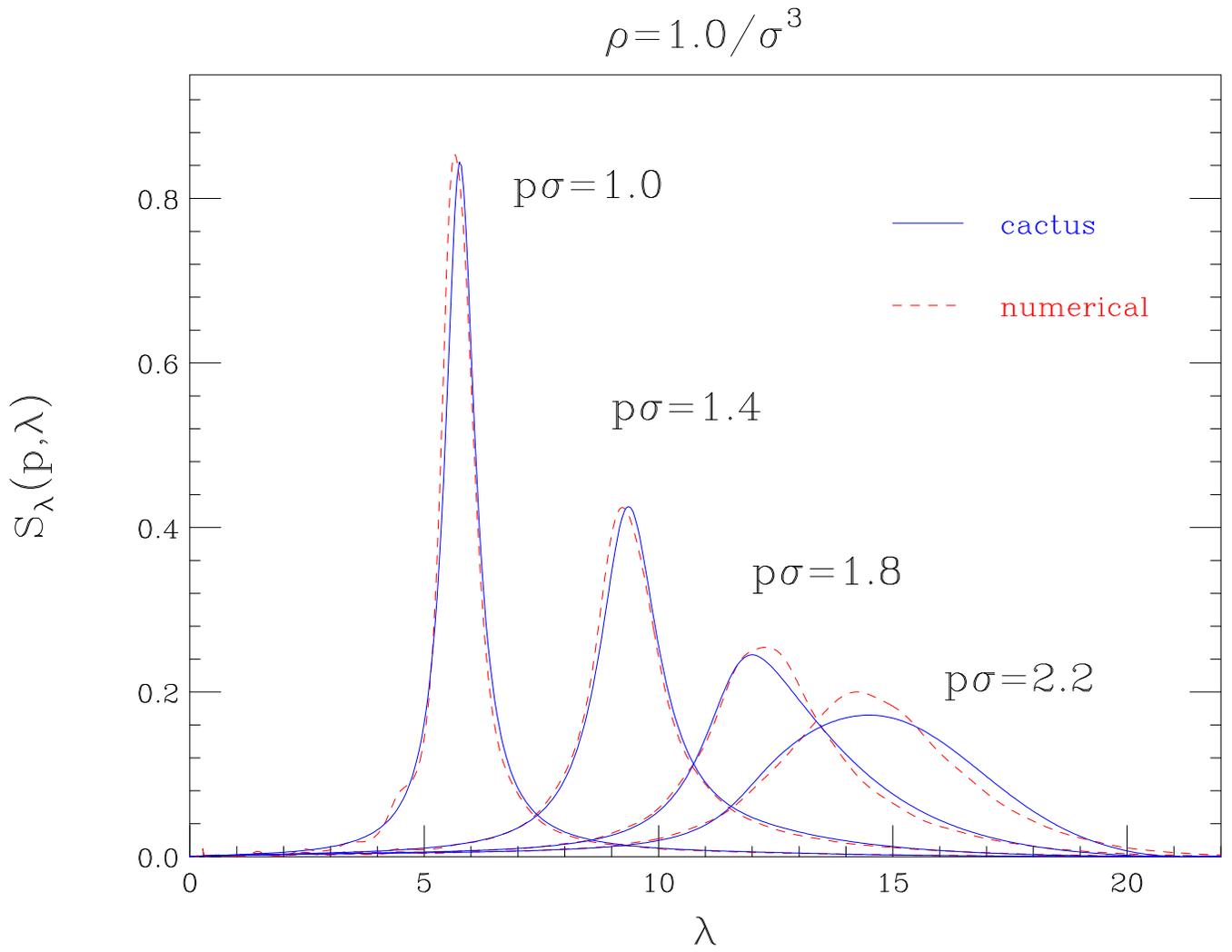, angle=90, width=\columnwidth}
\caption{Dynamical structure factor in the eigenvalues domain for 
several values of the momentum at $\rho=1$.}
\label{structure-factor}
\end{figure}

We have numerically solved equation~(\ref{CACTUS}) to find the
self-energy for several values of $\rho$ with the Gaussian choice
(\ref{GAUSSIANA}), thus obtaining the structure factor and the density
of states, in the eigenvalue domain.  In figure
(\ref{structure-factor}) we show $S^{(1)}(p,\lambda)$ for several
values of the momentum as obtained from equation (\ref{CACTUS}) for
$\rho=1$. We also plot the structure factor obtained numerically by
the method of moments (Turchi {\em et al.} 1982, Benoit 1989, Benoit
{\em et al.} 1992) (for lower momenta, the comparison with the method
of moments cannot be done due to finite volume effects, Martin-Mayor
{\em et al.} 2001).  A very good agreement with the numerical data is
achieved.  Note the absence of the secondary peak in this model (see
below).

\begin{figure}
\epsfig{file=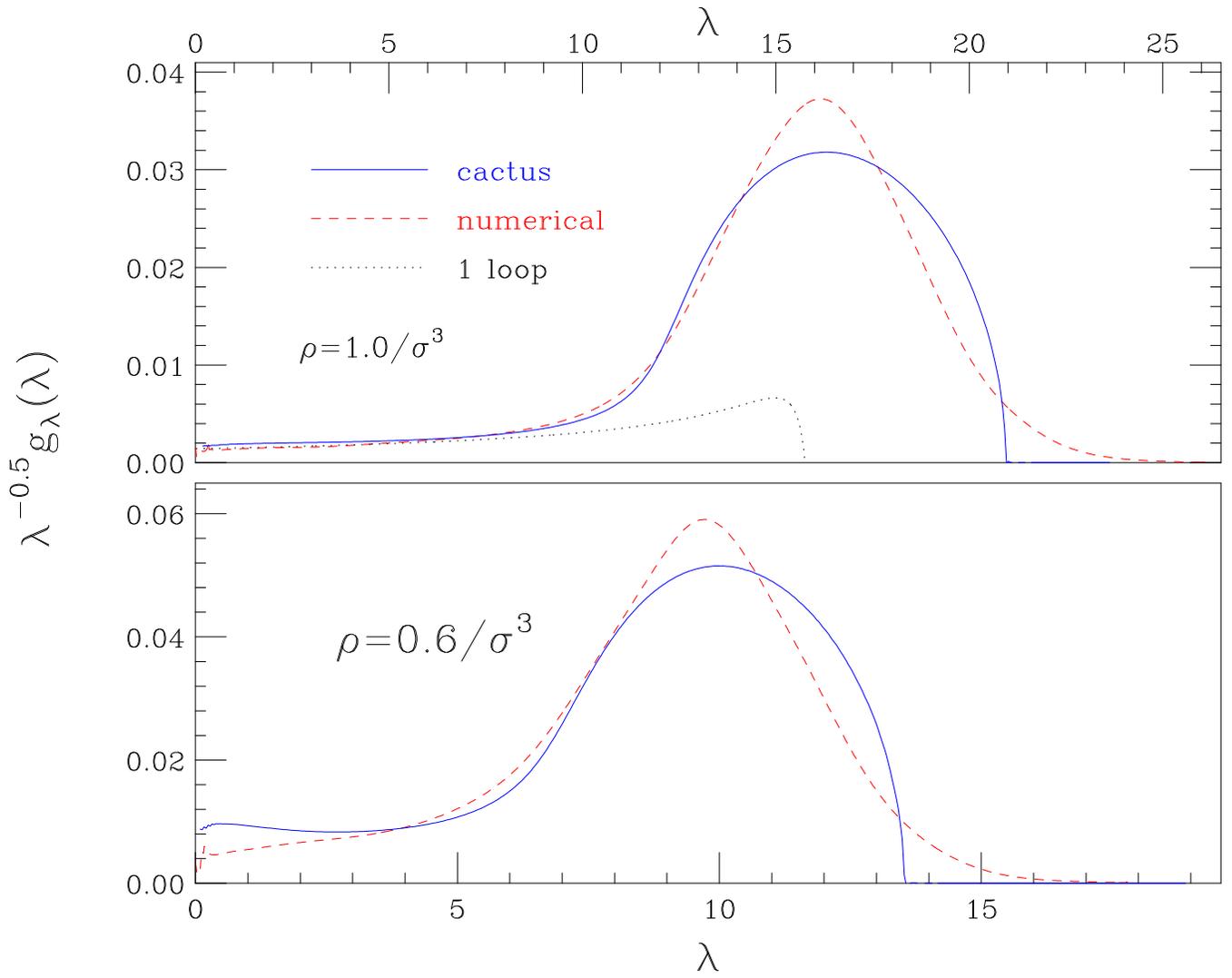, angle=90, width=\columnwidth}
\caption{Top: DOS in the eigenvalues domain divided by $\sqrt{\lambda}$ (Debye
behavior) for $\rho=1$. Bottom: DOS for $\rho=0.6$.}
\label{DOS}
\end{figure}

We found that for densities down to $\rho \approx 0.6$, the agreement
is good for all momenta, in fact good results are also obtained for
the density of states (see figure (\ref{DOS})). For lower densities,
the DOS starts to deviate from a Debye behavior at small $\lambda$, in
contradiction with the numerical results. Moreover, below $\rho
\approx 0.31$ the approximation gives a DOS with an unphysical
negative spectrum.  The negative spectrum develops continuously as a
function of $\rho$.  Also included in figure~(\ref{DOS}) is the
one-loop result. Notice that even for $\rho=1$, the cactus resummation
fails to reproduce the exponential decay of the density of states.
This is not unexpected, however, due to the non perturbative nature
(in $1/\rho$) of this tail (Zee and Affleck, 2000).

It is interesting to look at the small $\lambda$ limit of
$g_\lambda(\lambda)$. As seen in figure~(\ref{DOS}) (top) the behavior
is very nearly Debye (i.e.\ $\sqrt{\lambda}$) for small $\lambda$. An
excess of states relative to the Debye case develops for higher
$\lambda$, but in a region of eigenvalues well beyond the linear
dispersion regime (see bottom of figure (\ref{structure-factor})). Thus
this peak, similar to that found in (Schirmacher {\em et al.} 1998), 
should not be regarded as a Boson peak.

\begin{figure}
\epsfig{file=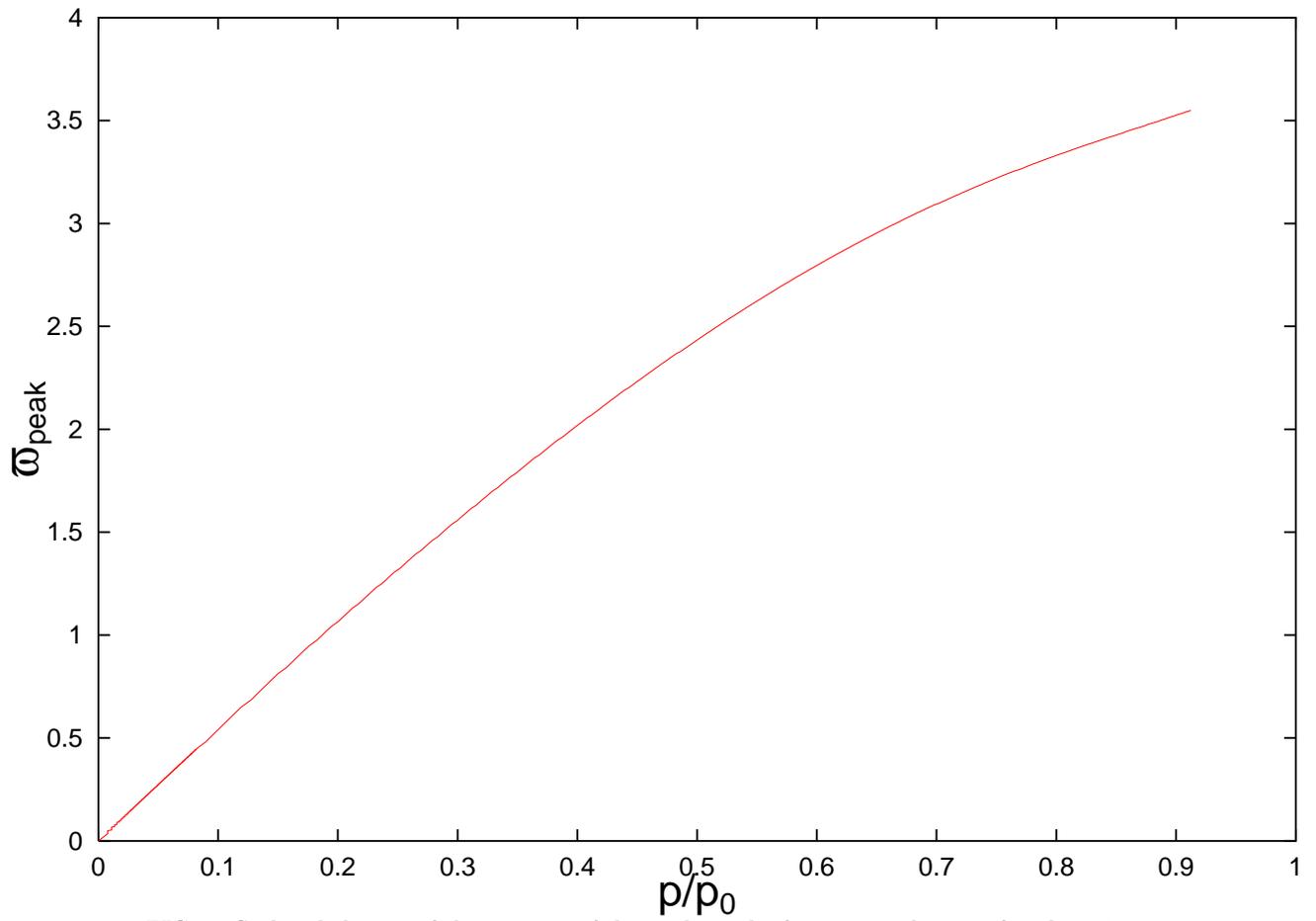, angle=270, width=\columnwidth}
\caption{Scaling behavior of the position of the peak, in the frequencies 
domain, for $rho=1$.}
\label{PICCO}
\end{figure}

\begin{figure}
\epsfig{file=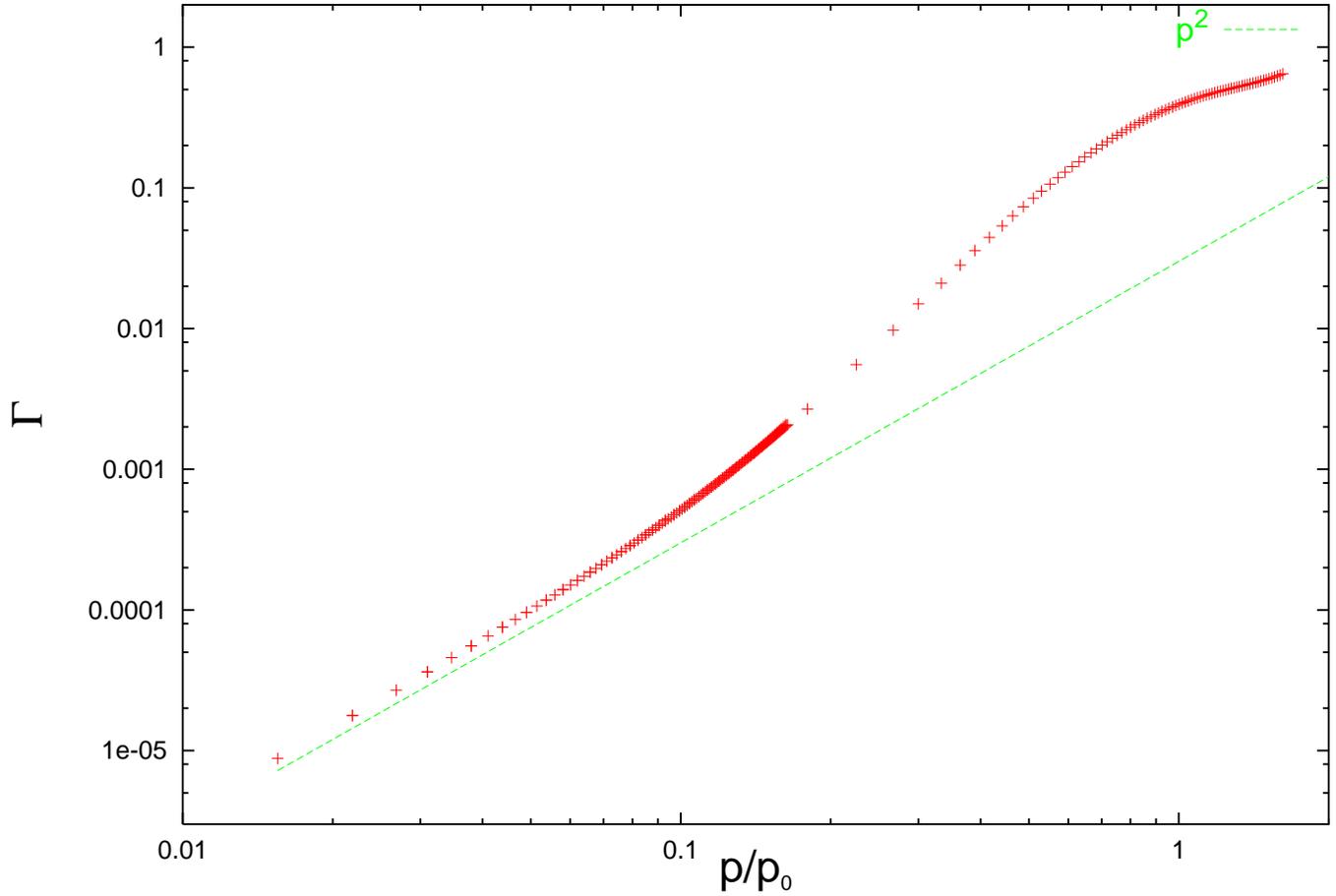, angle=270, width=\columnwidth}
\caption{Scaling behavior of the peak width at a function of $p$, in the 
frequencies domain, for $\rho =1$.}
\label{scaling}
\end{figure}

Finally, let us look at the scaling of the position and the width of
the peak in the frequency domain.  In figure~(\ref{PICCO}) we show the
frequency corresponding to the Brillouin peak as a function of the
external momentum $p$. Let us note that a nearly linear dependance,
(pointing likely to a propagating excitation) persists up to $p/p_0
\sim 0.6-0.7$.  In figure~(\ref{scaling}) instead we plot $\Gamma(p)$,
obtained by means of (\ref{GAMMA}).  As expected, the $p^2$ scaling is
found for very small momenta, which crosses over to a region where a
simple law as (\ref{BROAD}) is not suitable to describe the real
beahviour of the system.

Note that the region where the $p^2$ scaling is actually found, i.e.
$p/p_0 < 0.1$ is quite different from the region explored by X-rays
and neutrons scattering experiments, which rather spans the momenta
$0.1 < p/p_0 <0.5$, It is worthwile to note that the same conclusion
can be drawn from the results of G\"otze and Mayr using MCT for hard
spheres (G\"otze \& Mayr 2000).

\section{Conclusions and outlooks}


In summary, we have applied the random-matrix approach to the study of
the high-frequency excitations of glassy systems. We have presented a
resumation scheme (Grigera {\em et al.} 2001) that greatly enhances
the predictive power of the $1/\rho$ expansion prevoisuly obtained
(Martin-Mayor {\em et al.} 2001).  We have compared our analytical
calculation with numerical results, finding that for not too low
densities the only failure of the resumation scheme is its inability
to repoduce the exponential decay of the density of states.

Our calculations have been performed choosing a gaussian force, in
order to modelize the interparticle force {\em dressed} with the pair
correlation function, and only collinear displacements have been
considered.  We believe that our results are relevant for realistic
glasses, at least in the regime $p/p_0 < 1$.  The basic equation
(\ref{CACTUS}) shows that, in the general case, the position of the
spectral peak is linear at small momenta, and the numerical solution
of the integral equation in the Gaussian case shows that the linearity
persists up to $p/p_0 \sim 0.6-0.7$.

Moreover, the width of the spectral peak turns out to be proportional
to $p^2$ in the limit of $p\to 0$ (see
equation~(\ref{sigma-infinito})) irrespective of the potential
function $f(x)$.  Interestingly, in the Gaussian case the scaling law
holds for a momentum range one order of magnitude smaller than the
experimental one.  At larger momenta, the law is more complicated (see
figure (\ref{scaling})). Being this result quite similar to MCT result
for an hard-spheres system (G\"otze and Mayr 2000), we believe that it
could have a certain degree of univerality.

This model lacks an important feature of the experimental spectra,
namely the secondary peak of the $S(q,\omega)$ (which some authors
have interpreted as the Boson peak, see Horbach {\em et al.} 1998).
We believe this to be related to two important ingredients missing in
our model: the vectorial nature of the vibration, and a detailed
consideration of the particle correlations. The latter should not
affect our results for momenta much smaller than the first maximum of
the static structure factor, $p_0$. Since the secondary peak appears
for $p\sim p_0$, we should not really expect to describe it: in this
momentum range the dispersion relation is no longer monotonic in real
glasses. Furthermore, Dell'Anna {\em et al.} 1998 have suggested that
transversal excitations may play a prominent role in the raising of
the secondary-peak.  Our approach can be extended to include
transverse displacements, though, and work in this direction is in
progress.

VMM was partly supported by CICyT AEN99-0990, AEN97-1693 and
M.E.C. TSG was supported in part by CONICET (Argentina).

\section{References}
\noindent
Allen P.\ B. \ ,Feldman J.\ L.\,Fabian J.\ and Wooten F.\, 1999
{\em Phil. Mag. B} {\bf 79} \\ 
Angelani L.\, Di Leonardo R.\, Ruocco G.\, Scala A.\, Sciortino F.\, 2000
{\em Phys. Rev. Lett.} 85, 5356 \\
Benassi P.\, Krisch M.\,Masciovecchio C.\,Mazzacurati V.\, Monaco G.\, 
Ruocco G.\,Sette F.\, and Verbeni R.\, 1996
{\em Phys.\ Rev.\ Lett.}\ {\bf 77,} 3835 \\
Benoit C.\, 1989 {\em J. Phys.: Condens. Matter} {\bf 1,} 335 \\ 
Biroli G.\ and Monasson R.\, 1999 {\em J. Phys. A: Math. Gen.} {\bf 32,}
L255 \\
Buchenau U.\, Prager M.\, Nücker N.\, Dianoux A.\ J.\, Ahmad N.\, and 
Phillips W.\ A.\, 1986 {\em Phys.\ Rev.\ B} {\bf 34,} 5665 \\
Broderix K.\, Bhattacharya K.\ K.\, Cavagna A.\, Zippelius A.\, 
Giardina I.\, 2000 {\em Phys. Rev. Lett.} 85, 5360 \\
Cavagna A.\, Giardina I.\, Parisi G.\, 1999 {\em Phys. Rev. Lett.} 
{\bf 83,} 108 \\
Dell'Anna R.\, Ruocco G.\, Sampoli M.\ and Viliani G.\, 1998
{\em Phys.\ Rev.\ Lett.} {\bf 80,} 1236 \\
Ducastelle F. and Treglia G.\, 1982 {\em J. Phys. C} {\bf 15,} 2891 \\
Elliot S.\ R.\, 1983 {\em Physics of amorphous materials} 
(England: Longman) \\
Feldman J.\ L.\, Allen P.\ B.\ and  Bickham S.\ R.\, 1999 
{\em Phys.\ Rev.\ B} {\bf 59,} 3551  \\
Fioretto D.\, Buchenau U.\, Comez L.\, Sokolov A.\, 
Masciovecchio C.\, Mermet A.\, Ruocco G.\, Sette F.\,
Willner L.\, Frick B.\, Richter D.\, and Verdini L.\, 1999
{\em Phys.\ Rev.\ E} {\bf 59,} 4470  \\
Foret M.\, Courtens E.\, Vacher R.\,
Suck J.\ B.\, 1996 {\em Phys.\ Rev.\ Lett.} {\bf 77,} 3831 \\
G\"{o}tze W.\ and Mayr M.\ R.\, 2000 {\em Phys. Rev. E} {\bf 61,} 587 \\
Grigera T.S., Mart\'{\i}n-Mayor V., Parisi G. and Verrocchio P. 2001
cond-mat/0102230  \\
Horbach J.\, Kob W.\, Binder K.\, 1998 
{\em  J. Non-Cryst solids} {\bf 235,} 320 \\ 
Horbach J.\, Kob W.\, Binder K.\, 1999
{\em J.\ Phys.\ Chem.\ B} {\bf 103,} 4104 \\
Kantelhardt J.\ W.\, Russ S.\, Bunde A.\,
cond-mat/0012392; {\em Phys. Rev. B} (in press) \\ 
Keyes T.\, 1997 {\em J. Chem. Phys.} {\bf 101,} 2921  \\
Mart\'\i{}n-Mayor V.\, Parisi G.\, Verrocchio P.\, 2000 
{\em Phys.\ Rev.\ E} {\bf 62,} 2373  \\
Mart\'\i{}n-Mayor V.\, M\`ezard M.\, Parisi G.\, Verrocchio P.\, 2001 
{\em J. Chem. Phys.} (in press) \\
Masciovecchio C.\, Ruocco G.\, Sette F.\, Krisch M.\, 
Verbeni R.\, Bergmann U.\, and Soltwisch M.\, 1996
{\em Phys.\ Rev.\ Lett.} {\bf 76,} 3356  \\
Masciovecchio C.\, Ruocco G.\, Sette F.\, Benassi P.\, 
Cunsolo A.\, Krisch M.\, Mazzacurati V.\, Mermet A.\, 
Monaco G.\, and Verbeni R.\, 1997
{\em Phys.\ Rev.\ B} {\bf 55,} 8049 \\
Masciovecchio C.\, Monaco G.\, Ruocco G.\, Sette F.\,
Cunsolo A.\, Krisch M.\, Mermet A.\, Soltwisch M.\, and Verbeni R.\, 1998  
{\em Phys.\ Rev.\ Lett.} {\bf 80,} 544  \\
Mazzacurati V., Ruocco G.\ and Sampoli M., 1996, 
{\em Europhys. Lett.}  {\bf 34,} 681 \\
Metha M. L.\, 1991 {\em Random matrices}, (Academic Press) \\
M\`ezard M.\, Parisi G.\, Zee A.\, 1999 {\em Nucl. Phys.} 
{\bf B559,} 689  \\
Monaco G.\, Masciovecchio C.\, Ruocco G.\, Sette F.\, 1998 
{\em Phys.\ Rev.\ Lett.} {\bf 80,} 2161 \\
Montagna M.\, Ruocco G.\, Viliani G.\, Dell'Anna R.\, 
Di Leonardo R.\, Dusi R.\, Monaco G.\, Sampoli M.\, and Scopigno T., 1999 
{\em Phys. Rev. Lett.} {\bf 83}, 3450 \\
Parisi G.\, 1994 {\em Field-Theory, Disorder and Simulations}, World 
Scientific.
\\
Ribeiro M.\ C.\ C.\, Wilson M.\ and Madden P.\ A.\, 1998
{\em J.\ Chem.\ Phys.} {\bf 108,} 9027  \\
Ruocco G.\, Sette F.\, Di Leonardo R.\, Fioretto D.\, 
Krisch M.\, Lorenzen M.\, Masciovecchio C.\, Monaco G.\, 
Pignon F.\ and Scopigno T.\, 1999 
{\em Phys.\ Rev.\ Lett.\ } {\bf 83,} 5583 \\
Ruocco G.\, Sette F.\, Di Leonardo R.\, Monaco G.\, 
Sampoli M.\, Scopigno T.\, and Viliani G.\, 2000 
{\em Phys.\ Rev.\ Lett.} {\bf 84,} 5788 \\
Sampoli M.\, Benassi P.\, Dell'Anna R.\, Mazzacurati V.\ 
and Ruocco G.\ 1998, {\em Phil. Mag. B} {\bf 77,} 473 \\
Schirmacher W.\, Diezemann G.\ and Ganter C.\, 1998 
{\em Phys.\ Rev.\ Lett.\ } {\bf 81,} 136  \\
Sette F.\, Krisch M.\, Masciovecchio C.\, Ruocco G.\ and Monaco G.\,1998
{\em Science} {\bf 280,} 1550 \\
Sokolov A.\ P.\, Buchenau U.\, Richter D, Masciovecchio C.\, 
Sette F.\, Mermet A.\, Fioretto D.\, Ruocco G.\, Willner L.\
and Frick B.\, 1999 {\em Phys.\ Rev.\ E} {\bf 60,} 2464 \\
Taraskin S.\ N.\ and Elliot S.\ R.\, 1999 {\em Phys.\ Rev.\ B} 
{\bf 59,} 8572  \\
Taraskin S.\ N.\,Loh Y. L.\,Natarajan G.\ and Elliott S.\ R.\, 2001
 {\em Phys.\ Rev.\ Lett.\ } {\bf 86,} 1255 \\
Turchi P.\, Benoit C.\, Royer E.\ and Poussigue G.\, 1992
{\em J. Phys.: Condens. Matter} {\bf 4,} 3125  \\
Wan Y.\ and Stratt R.\, 1994 {\em J. Chem. Phys} {\bf 100,} 5123 \\
Wu T.\ M.\ and Loring R.\ F.\, 1992 {\em J. Chem. Phys.} {\bf 97,} 8368 \\
Zee A.\ and Affleck I.\, 2000
{\em J. phys: cond. mat.}  {\bf 12,} 8863 \\

\newpage

\newpage

\newpage

\newpage

\newpage

\newpage

\newpage

\end{document}